\documentclass[12pt,twoside]{article}
\usepackage{amssymb,amsmath}
 
\newcommand{\ket}[1]{\left|#1\right\rangle}
\newcommand{\bra}[1]{\left\langle#1\right|}
\newcommand{\braket}[2]{\left\langle{#1}\mkern-2mu
\mid\mkern-2mu{#2}
\right\rangle}
\newcommand{\numEq}[2]{\begin{equation}
   \label{eq:#1}
   #2\end{equation}}
\newcommand{\refEq}[1]{(\ref{eq:#1})}
\newcommand{\citeRef}[1]{\thinspace\cite{ref:#1}}
\newcommand{\eps}{\epsilon}
\newcommand{\Prob}{\mathop{\rm Prob}\nolimits}

\let\epsilon\varepsilon
\let\hat\widehat

 \newcommand{\Cases}[1]{\left\{ 
  \begin{array}{r@{\ ,}@{\quad}c}
  #1
  \end{array}\right.}

\numberwithin{equation}{section}

\begin{document}

\title{\mbox{}\vspace{-1.3in}\mbox{}\\Invariant Quantum Algorithms for\\
 Insertion into an Ordered List}
\markboth{Invariant Quantum Algorithms for Insertion into an Ordered
List}{E. Farhi, J. Goldstone, S. Gutmann, and M. Sipser}

\author{Edward Farhi, Jeffrey Goldstone\thanks{farhi@mit.edu ;   
goldstone@mitlns.mit.edu}\\[-0.5ex] \small
 Center for Theoretical Physics\\[-0.7ex] \small 
 Massachusetts Institute of Technology\\[-0.7ex] \small 
 Cambridge, MA  02139\\[1.5ex]
Sam Gutmann\thanks{sgutm@nuhub.neu.edu}\\[-0.5ex] \small
 Department of Mathematics\\[-0.7ex] \small
  Northeastern University\\[-0.7ex] \small 
 Boston, MA 02115\\[1.5ex]
Michael Sipser\thanks{sipser@math.mit.edu
\protect\newline This work was
supported in part by The Department of Energy under
cooperative agreement\protect\\ DE--FC02--94ER40818 and by the National
Science Foundation under grant NSF 95--03322 CCR.}\\[-0.5ex]\small
 Department of Mathematics\\[-0.7ex] \small 
 Massachusetts Institute of Technology\\[-0.7ex] \small
 Cambridge, MA  02139\\[2ex]\small
MIT CTP \# 2815\\[1ex]}

\date{\small January 19, 1999}

\maketitle
\thispagestyle{empty}

\begin{abstract}\noindent\small
We consider the problem of inserting one item into 
a list of $N-1$ ordered items. We previously 
showed that no quantum algorithm could solve this problem in 
fewer than $\log N/(2 \log \log N)$ queries, for $N$ large.
We transform the problem into a ``translationally
invariant'' problem and restrict attention to invariant 
algorithms. We construct the ``greedy'' invariant 
algorithm and show numerically that it outperforms  
the best classical algorithm for various~$N$. We also
find invariant algorithms that succeed exactly in
fewer queries than is classically possible, and 
iterating one of them shows that the insertion 
problem can be solved in fewer than 0.53~$\log N$ quantum 
queries for large $N$ (where $\log N$ is the classical lower 
bound). We don't know whether a $o(\log N)$ algorithm exists.

\end{abstract}

\newpage

\section{Introduction}\label{sec:1}

We consider the problem of inserting a new item into an
ordered list of $N-1$ items. A single classical query
consists of comparing the new item with any chosen item on
the list to see if the new item comes before or after the
chosen item. Classically, the best algorithm for
determining the point of insertion is binary search, which
uses $\lceil \log_2 N \rceil$ queries. In~\citeRef1 we showed that
quantum mechanically, for large~$N$, an algorithm that succeeds
after $k$ quantum queries must have 
 \numEq{1}{ k >  \frac{\log_2 N}{2\log_2 \log_2 N} \ .
 }
The same bound holds for algorithms that succeed with probability
$\eps>0$ (independent of~$N$).

 In this paper 
 we transform
 the insertion problem into an equivalent ``translationally
 invariant'' problem and restrict our attention to translationally
 invariant algorithms.  In the next section we spell out what we
mean by a translationally invariant algorithm. We derive a lower
bound on the number of quantum queries needed for a successful
translationally invariant algorithm. This bound turns out to
coincide with
\refEq1, which suggests to us that the best algorithm may in fact
be translationally invariant. 

In Section~\ref{sec:3} we construct the
greedy translationally invariant algorithm for the insertion
problem. By a ``greedy'' algorithm we mean an algorithm in which
each step is chosen to maximize the probability of success
after all preceding steps have been chosen. We present some numerical
results for the greedy algorithm. For example, if $N=2048$, after
5~quantum queries the probability of success is 0.9939 compared
to the best possible classical probability of 1/64. However, we have
not been able to analyze the large~$N$ behavior of the greedy
algorithm.

The greedy algorithm can achieve a high probability of success but
is not exact (``Exact'' means that the correct answer is guaranteed.)
In Section~\ref{sec:4} we present a method for
exploring whether an exact $k$-quantum-query translationally
invariant algorithm exists for a given~$N$. Using this method we
find a 2-query algorithm for $N=6$. A self-contained presentation of
this algorithm is given at the end of Section~\ref{sec:4}. Furthermore, we
find that no 2-query translationally invariant algorithm exists for $N\ge7$.
With 3 quantum queries we can construct a translationally invariant
algorithm for $N=52$ but we do not know how large a value of $N$ can be
attained with $k=3$. 

Starting with a $k$-quantum-query algorithm that exactly solves
the insertion problem for some~$M$, one can solve the insertion
problem for $N=M^h$ with $hk$ quantum queries for any
positive integer~$h$. To do this first pick out $M-1$ items, equally
spaced in the list of $M^h - 1$ items. Running the $k$-quantum-query algorithm determines the point of insertion to lie in a range
of $M^{h-1}$  items. Iterate this procedure a total of $h$ times to
exactly determine the point of insertion in the original list of $M^h
-1 $ items. 
Note that the overall algorithm with $hk$ queries is not
translationally invariant although the $k$-query subroutine is.

The result of the previous paragraph and our  exact $N=52$ in
$k=3$ algorithm (see Section~\ref{sec:4}) shows that one can
construct a quantum algorithm for solving the insertion problem
with $N-1$ items where the number of queries grows like 
\numEq{2.1}{
\Bigl( \frac3{\log_2 52}\Bigr) \log_2 N\ .
}
Further exploration of the methods in Section~\ref{sec:4} will
certainly lead to a better constant than $3/\log_2 52$ and
perhaps even an $o(\log N)$ algorithm. 

Recently, R\"ohrig\citeRef{2} published an algorithm that
uses an average of $(3/4)\log_2 N + O(1)$ queries to solve
the insert problem with probability $1/2$.  This is not
attainable classically, but iterating the algorithm to improve the
$1/2$ probability involves more queries than are required to solve
the insertion problem classically. 

Our results carry over immediately to sorting. Classically, in
the comparison model, $n$ items can be sorted in
$n\log_2 n +O(n)$ queries using binary-search insertion
for each $N=2,3,\dots,n$. Using our exact quantum insertion
algorithm as a subroutine, the number of required queries can be
cut by a constant factor, beating the classical lower bound of 
$n\log_2 n$.

\section{Translationally Invariant Algorithms}\label{sec:2}

The classical problem of inserting one item into an ordered list 
of $N-1$ items is equivalent to the following oracular problem:
Consider the $N$ functions $f_j$ defined on the set $\{0,1,\dots,
N-1\}$ by 
\numEq{3}{
f_j(x) =\Cases{
-1 &  x<j     \\
+1 &  x\ge j }
}
for $j=0,1,\dots,N-1$. A query consists of giving the oracle a value
of $x$ with the oracle returning $f_j(x)$ for some fixed but
unknown~$j$. The problem is to determine~$j$. (Note that 
$f_j(N-1)= +1$ for all $j$, so querying the oracle at $x=N-1$ is of no
help. However, it is convenient for us to include this value of~$x$.)

In order to construct our quantum algorithms we double the
domain of the functions $f_j$ and define
\numEq{4}{
F_j(x) =\Cases{
f_j(x) &  0\le x \le N-1    \\
-f_j(x-N) &  N\le x \le 2N-1\rlap{\ . }}
}
The problem is still to determine the value of~$j$.  Counting
queries of $F_j$ is equivalent to counting queries of~$f_j$. 
 Doubling the
domain of the functions is of no help classically but is of use to us
in the quantum setting. 
Note that $F_{j+1}(x) = F_j(x-1)$ for $j=0,1,\dots,N-2$ if we make
the identification that $x=-1$ is $x=2N-1$. In this sense the $F_j$'s
are translates of each other.

We work
in a Hilbert space of dimension
$2N$ with basis vectors $\ket x$ with $x=0,1,\dots,2N-1$. A
quantum query is an application of the unitary operator
\numEq{5}{
\hat F_j \ket x = F_j(x) \ket x
}
when the oracle holds the function~$F_j$. 
(The workbits necessary for constructing \refEq5 have been
suppressed.) A $k$-query quantum
algorithm starts in a state $\ket s$ and alternately applies $\hat
F_j$ and $j$-independent unitary operators $V_\ell$ to produce the
state
\numEq{6}{
V_k \hat F_j V_{k-1} \cdots V_1 \hat F_j \ket s\ .
}
(In our algorithms, all of the operators in \refEq6 act as the identity
in the suppressed work space.)  An algorithm succeeds if the states
in
\refEq6 are an orthogonal set for $j=0, 1,\dots, N-1$. 
Because the last unitary operator, $V_k$, is at our disposal we are
free to choose the orthogonal states of a successful algorithm to be
any orthogonal set. Corresponding to $F_j$, we choose
\numEq{7}{
 \ket {j+} = \frac1{\sqrt2} \bigl(\ket j +
                \ket{j+N}\bigr)
\quad\mbox{for $j=0,1,\dots,N-1$}}
to be the target state of a successful $k$-query algorithm
for~$k$~even and 
\numEq{8}{
 \ket {j-} = \frac1{\sqrt2} \bigl(\ket j -
\ket{j+N}\bigr)\quad\mbox{for
$j=0,1,\dots,N-1$}}
to be the target state of a successful $k$-query algorithm for
$k$~odd. (We defer the explanation for the odd/even distinction
until later.)

We now note that the $\hat F_j$ are translates of each other in the
following sense. Let the translation operator~$T$ be defined by
\begin{align}
&T\ket x = \ket{x+1}\quad
\mbox{for $x=0,1,\dots,2N-2$}\nonumber\\ 
&T\ket{2N-1} = \ket
0\ .
\label{eq:9}
\end{align}
Then we have 
\numEq{10}{
 T\hat F_j T^{-1} = \hat F_{j+1}
}
for $j=0,1,\dots,N-2$ and equivalently
\numEq{11}{
 T^j \hat F_0 T^{-j} = \hat F_j
}
for $j=1,2,\dots,N-1$. Furthermore
\numEq{12}{
 T^j \ket{0\pm} = \ket{j\pm}
}
for $j=1,2,\dots,N-1$. 

Suppose we pick the starting state of our algorithm to be 
\numEq{13}{
 \ket s = \frac1{\sqrt{2N}} \sum_{x=0}^{2N-1}\ket x
}
which is translationally invariant, that is,
\numEq{14}{
 T \ket s = \ket s \ .
}
Furthermore, suppose we limit ourselves to translationally
invariant unitary operators $V_\ell$, that is, we require
\numEq{15}{
 T V_\ell T^{-1} = V_\ell\quad \mbox{for $\ell=1,2,\dots,k$.}
}
Then if a $k$-query algorithm succeeds for $j=0$, that is,
\numEq{16}{\begin{align}
 \ket{0+} &= V_k \hat F_0 V_{k-1} \cdots V_1 \hat F_0 \ket
s\quad \mbox{when $k$ is even, or}\\
\ket{0-} &= V_k \hat F_0 V_{k-1} \cdots V_1 \hat F_0 \ket
s\quad \mbox{when $k$ is odd}\nonumber
\end{align}}
then because of \refEq{11},  \refEq{12}, \refEq{14}, and \refEq{15}
it follows that 
\numEq{17}{\begin{align}
 \ket{j+} &= V_k \hat F_j V_{k-1} \cdots V_1 \hat F_j \ket s\quad
\mbox{when $k$ is even, or}\\
\ket{j-} &= V_k \hat F_j V_{k-1} \cdots V_1 \hat F_j \ket s\quad
\mbox{when $k$ is odd.}
\nonumber
\end{align} }
A clear advantage of this translationally invariant \emph{ansatz} is
that finding a set of $V$'s which makes the single $j$-independent
condition \refEq{16} hold guarantees that the algorithm succeeds
for all~$j$. 

To understand which operators $V$ are translationally invariant,
that is, satisfy $TVT^{-1}=V$, we work in the momentum basis
\numEq{18}{
 \ket{\bf p} = \frac1{\sqrt{2N}}\sum_{x=0}^{2N-1} e^{i{\bf p} x\,
\pi/N}
\ket x\quad\mbox{for ${\bf p}=0,1,\dots, 2N-1$}
}
for which
\numEq{18a}{
 \ket{x} = \frac1{\sqrt{2N}}\sum_{{\bf p}=0}^{2N-1} e^{-i{\bf p} x\,
\pi/N}
\ket {\bf p}\quad\mbox{for $x=0,1,\dots, 2N-1$.}
}
Kets with boldfaced labels always denote momentum basis vectors.
Note that
\numEq{19}{
 T\ket{\bf p} =  e^{-i{\bf p} \, \pi/N}
\ket {\bf p}
}
so we see that $T$ is diagonal in the momentum basis. Thus if
$V_\ell$ is diagonal in the momentum basis, that is, 
\numEq{20}{
  V_\ell \ket{\bf p} = e^{i\alpha_\ell({\bf p})}
\ket {\bf p}
}
where $\alpha_\ell({\bf p})$ is real, then each $V_\ell$ is both
unitary and translationally invariant. 

Constructing a successful $k$-query translationally invariant
algorithm is equivalent to finding phases $\alpha_\ell({\bf p})$, for
$\ell=1,2,\dots,k$ to make \refEq{16} hold. Because of \refEq1, for
a given~$N$, we know that this cannot be done if $k$ is too small. 
Strategies for choosing  the phases
$\alpha_\ell({\bf p})$ for the greedy algorithm and for exactly
successful algorithms are the subjects of the next two sections.

Because the translationally invariant \emph{ansatz} has led to the
momentum basis, all the elements of \refEq{16} are best expressed
in the momentum basis. The $V_\ell$'s are defined in the
momentum basis by~\refEq{20}. By~\refEq{13} and~\refEq{18}
we have 
\numEq{21}{
\ket s = \ket{\bf 0} \ .
}
(Recall that the boldface $\bf 0$ denotes the momentum basis
vector with $\bf p=0$.) By \refEq7, \refEq8, and \refEq{18a} we
have
\begin{align}\label{eq:22}
\ket {0+} &= \frac1{\sqrt2}(\ket0 + \ket N) = \frac1{\sqrt N}
\sum_{{\bf p}\;{\rm even}} 
\ket{\bf p}\\
\intertext{\noindent and}
\ket {0-} &=\frac1{\sqrt2}(\ket0 - \ket N) =  \frac1{\sqrt N}
\sum_{{\bf p}\;{\rm odd}} 
\ket{\bf p}\ .\nonumber
 \end{align}
(The nonboldfaced kets $\ket0$ and $\ket N$ are in the $\ket x$
basis.) We also need the matrix elements of $\hat F_0$ in the
momentum basis, 
\begin{align}
\bra{\bf p} \hat F_0 \ket{\bf q} &= 
\sum_{x=0}^{2N-1} \bra{\bf p} \hat F_0 \ket x \braket{x}{\bf q}
\label{eq:23}\\
&= \sum_{x=0}^{2N-1}  \braket{\bf p}{x}  F_0 (x) 
         \braket{x}{\bf q} \nonumber\\
&= \frac1{2N} \Bigl(\sum_{x=0}^{N-1} - \sum_{x=N}^{2N-1}\Bigr)
      e^{i \pi({\bf q}-{\bf p}) \, x/N}\ .\nonumber
\end{align}
So
\numEq{24}{
\bra{\bf p} \hat F_0 \ket{\bf q} = \Cases{
{\displaystyle \frac{i e^{-i \pi({\bf q}-{\bf p})/2N}}{N\sin \pi({\bf
q}-{\bf p})/2N}} & \mbox{${\bf q}-{\bf p}$ odd}\\[2.5ex]
0\hfil & \mbox{${\bf q}-{\bf p}$ even.}
}}

After $\ell$ queries, a translationally invariant algorithm produces
the state 
\numEq{25}{
\ket{\psi_\ell} = V_\ell \hat F_0 V_{\ell-1}\cdots V_1\hat F_0
\ket{\bf0}\ .
 }
Here and throughout, $k$ is the fixed total number of queries, and
$\ell$, with $1\le\ell\le k$, indexes a stage of the
algorithm. Expressed in the momentum basis, for
$\ell$ even, using
\refEq{20} we have 
\begin{align}
\ket{\psi_\ell} &= \sum_{{\bf p}_\ell\; {\rm even}}\;
\sum_{{\bf p}_{\ell-1}\; {\rm odd}}\cdots\sum_{{\bf p}_1\; {\rm
odd}}
\label{eq:26} \\
  &\qquad 
\ket{{\bf p}_\ell} e^{i\alpha_\ell({\bf p}_\ell)}
\bra{{\bf p}_\ell} \hat F_0 \ket{{\bf p}_{\ell-1}}
e^{i\alpha_{\ell-1}({\bf p}_{\ell-1})} \cdots
e^{i\alpha_1({\bf p}_1)}
\bra{{\bf p}_1} \hat F_0 \ket{{\bf 0}} \nonumber
\end{align}
where we need only include ${\bf p}_1$~odd, ${\bf
p}_2$~even,~etc.\ because of \refEq{24}. This means that at
each stage there are~$N$, not~$2N$, phases to choose. 

The goal of an algorithm is to produce the state $\ket{0+}$ after an
even number of queries (or  $\ket{0-}$ after an odd number). We
can judge how close to success we are at the $\ell$-th stage by
evaluating the overlap with $\ket{0+}$ (for $\ell$~even),
\numEq{27}{
\braket{0+}{\psi_\ell} = \frac1{\sqrt N} \sum_{{\bf p}\text{ even}}
\braket{{\bf p}}{\psi_\ell} 
}
by \refEq{22}.
For these translationally invariant algorithms,
the probability of success
if we stop at the $k$-th stage is the same whichever $F_j$ the oracle
holds and equals $\bigl|\braket{0+}{\psi_k}\bigr|^2$. To find a lower
bound on the number of queries required for success we note, using
\refEq{26} and \refEq{27} that
\begin{align}
\Bigl|\braket{0+}{\psi_\ell}\Bigr| \le \frac1{\sqrt N} 
&\sum_{{\bf p}_\ell\; {\rm even}}\;
\sum_{{\bf p}_{\ell-1}\; {\rm odd}}\cdots \sum_{{\bf p}_1\; {\rm
odd}}
\label{eq:28} \\
  &\quad
\Bigl| \bra{{\bf p}_\ell} \hat F_0 \ket{{\bf p}_{\ell-1}}
\bra{{\bf p}_{\ell-1}} \hat F_0 \ket{{\bf p}_{\ell-2}}
\cdots
\bra{{\bf p}_1} \hat F_0 \ket{{\bf 0}}\Bigr|\ . \nonumber
\end{align}
Because $\bra{\bf p} \hat F_0 \ket{\bf q}$ only depends on 
$({\bf p} -{\bf q})\!\!\mod 2N$ the righthand side of \refEq{28} consists of
$\ell$ identical factors and we have
\numEq{29}{
\Bigl|\braket{0+}{\psi_\ell}\Bigr| \le \frac1{\sqrt N} 
\Bigl[\sum_{{\bf p}\; {\rm odd}}
\Bigl|\bra{{\bf p}} \hat F_0 \ket{{\bf 0}}\Bigr|
\Bigr]^\ell\ .
}
By \refEq{24} we then have
\numEq{30}{
\Bigl|\braket{0+}{\psi_\ell}\Bigr| \le \frac1{\sqrt N} 
\Bigl[\frac1N \sum_{{\bf p}\; {\rm odd}}
\frac1{\sin\pi{\bf p}/2N}
\Bigr]^\ell\ .
}
(Recall that $0\le{\bf p}\le 2N-1$.) Approximating the sum,
we have 
\begin{align}
\frac1N \sum_{{\bf p}\; {\rm odd}} \frac1{\sin\pi{\bf p}/2N} &=
\frac4\pi \Bigl(1+\frac13+\dots+\frac1{N-1}
 \Bigr)\label{eq:31}\\
&\qquad{} + 
\frac2\pi \int_0^{\pi/2} {\rm d}\theta \Big(\frac1{\sin\theta}
-\frac1\theta \Bigr) + O(1/N)\nonumber\\[1.5ex]
&= \frac2\pi \Bigl(\ln N +\gamma +\ln\frac8\pi \Bigr) + O(1/N)\ .
\nonumber
\end{align}
(The approximation \refEq{31} is already correct at $N=3$ to
1~part in~1000.)
 A $k$-query algorithm that succeeds with
probability~$\eps$ must have
\numEq{32}{
\eps\le \Bigl|\braket{0+}{\psi_k}\Bigr|^2 \le \frac1N
\Bigl[\frac2\pi \ln N +O(1)
\Bigr]^{2k} 
}
which implies, for $N$ large, that 
\numEq{33}{
k > \frac{\ln N}{2\ln\ln N}\ .
}
Note that the ratio of the  righthand sides of the bounds \refEq1
and \refEq{33} converges to~1 as $N\to\infty$. Since the bound
\refEq{33} was derived under the assumption of translation
invariance, it is only a special case of the fully general
bound~\refEq1.

   The idea of invariance makes sense in other 
computing problems. For example, Grover's search 
problem~\cite{ref:3} is invariant under the group of 
permutations. Requiring an algorithm in this 
problem to be permutation invariant is extremely
restrictive. There is only one phase to choose for
each $V$, and it is easy to see that the choice of
$-1$ at each stage, which corresponds to Grover's
algorithm, is optimal.

\section{The Greedy Algorithm}\label{sec:3}

The state produced after $\ell$ queries of a translationally
invariant algorithm, given in \refEq{25}, can be related to the state
produced after $\ell-1$ queries by 
\numEq{3.1}{
\ket{\psi_\ell} = V_\ell \hat F_0 \ket{\psi_{\ell-1}}
}
where $\ket{\psi_0} = \ket{\bf 0}$.
We define the greedy algorithm inductively. Given
$\ket{\psi_{\ell-1}}$ we choose $V_\ell$ to maximize the overlap
of $\ket{\psi_\ell}$ with $\ket{0+}$ if $\ell$ is even or with
$\ket{0-}$ if $\ell$ is odd. At each stage the overlap increases and
hence the probability of success if we stop at the $k$th stage increases
with $k$.  As we will see below, the greedy algorithm is never perfect,
but we provide numerical evidence that it converges rapidly. For selected
values of $N$ up to 4096 we see that the greedy algorithm outperforms the
best classical algorithm.

We begin by showing how well the greedy algorithm does with one
query. In this case
\begin{align}
\ket{\psi_1} &= V_1 \hat F_0 \ket{\bf 0} \label{eq:3.2}\\
&= \sum_{{\bf p}\; {\rm odd}}
\ket{\bf p} e^{i\alpha_1({\bf p})}
\bra{\bf p} \hat F_0 \ket{\bf 0}\nonumber
\end{align}
where we have inserted a complete set, used \refEq{20} and
used the fact that $\bra{\bf p} \hat F_0 \ket{\bf 0}$ vanishes for
$\bf p$ even. By \refEq{22}
\numEq{3.3}{
\Bigl|\braket{0-}{\psi_1}\Bigr| =
 \frac1{\sqrt N} 
\Bigl| \sum_{{\bf p}\; {\rm odd}} e^{i\alpha_1({\bf p})}
\bra{\bf p} \hat F_0 \ket{\bf 0} \Bigr|\ .
}
Choosing $V_1$ is equivalent to choosing the phases
$\alpha_1({\bf p})$.  To maximize \refEq{3.3} we choose
$\alpha_1({\bf p})$ to make each term in the sum real and
positive. With this choice
\numEq{3.4}{
\Bigl|\braket{0-}{\psi_1}\Bigr| =
 \frac1{\sqrt N} 
\sum_{{\bf p}\; {\rm odd}} \Bigl| 
\bra{\bf p} \hat F_0 \ket{\bf 0} \Bigr|\ .
}
Using \refEq{24} we have
\numEq{3.5}{
\Bigl|\braket{0-}{\psi_1}\Bigr| =
 \frac1{N^{3/2}} 
\sum_{{\bf p}\; {\rm odd}} \frac1{\sin(\pi{\bf p}/2N)}\ . }
Approximating the sum, for $N$ large, as we did in \refEq{31},
gives
\numEq{3.6}{
\Bigl|\braket{0-}{\psi_1}\Bigr|^2 \sim
 \frac{4}{\pi^2 N}\Bigl[\ln N + \gamma + \ln \frac8\pi \Bigr]^2}
which is the probability of success after running a $1$-query
greedy algorithm. This beats the classically best possible, which
is~$2/N$. 

To see how the greedy algorithm works at the $\ell$-th stage
($\ell$~even, for example) first note that by \refEq{22} 
\begin{align}\label{eq:3.7} 
\Bigl| \braket{0+}{\psi_\ell} \Bigr| &= 
\frac1{\sqrt N} \Bigl| \sum_{{\bf p}\; {\rm
even}}\braket{\bf p}{\psi_\ell}  \Bigr|\\
& = \frac1{\sqrt N} \Bigl| \sum_{{\bf p}\; {\rm
even}}
e^{i\alpha_\ell({\bf p})}
\bra{\bf p} \hat F_0 \ket{\psi_{\ell-1}}\Bigr| \nonumber
\end{align}
by \refEq{3.1} and \refEq{20}. To maximize \refEq{3.7} we choose
the phases
$\alpha_\ell({\bf p})$ to make each term in the sum real and
nonnegative, that is, each $\braket{\bf p}{\psi_\ell}$ is real and
nonnegative. Now 
\numEq{3.8}{
\braket{\bf p}{\psi_\ell} =
 e^{i\alpha_\ell({\bf p})} \sum_{{\bf q}\; {\rm odd}}
\bra{\bf p} \hat F_0 \ket{\bf q} \braket{\bf q}{\psi_{\ell-1}}
}
and by the choice of $\alpha_\ell({\bf p})$ and \refEq{24}
\begin{align}\label{eq:3.9}
\braket{\bf p}{\psi_\ell} &=
\frac1N \Bigl| \sum_{{\bf q}\; {\rm odd}}
\frac{e^{i\pi({\bf p}-{\bf q})/2N}}{\sin \pi({\bf p}-{\bf q})/2N}
\braket{\bf q}{\psi_{\ell-1}}  \Bigr| \\
&= \frac1N \Bigl| \sum_{{\bf q}\; {\rm odd}}
\Bigl(\cot \bigl(\pi({\bf p}-{\bf q})/2N\bigr) + i\Bigr)
\braket{\bf q}{\psi_{\ell-1}}  \Bigr|\ .\nonumber 
\end{align}
This last formula, together with its virtually identical $\ell$-odd
analogue,  explicitly determines
$\ket{\psi_\ell}$ from $\ket{\psi_{\ell-1}}$, providing a complete
description of the greedy algorithm. The choice of $k$, the number of
queries before stopping and measuring, depends on the probability of
success desired.

The probability of success after $\ell$~queries is 
\numEq{3.10}{
\Prob(\ell) = \Bigl| \braket{0+}{\psi_\ell} \Bigr|^2 = 
\frac1N \Bigl| \sum_{{\bf p}\; {\rm even}}\braket{\bf p}{\psi_\ell} 
\Bigr|^2\ . }
Now we can rewrite \refEq{3.9} as 
\numEq{3.11}{
\braket{\bf p}{\psi_\ell} =
\frac1N \Bigl| \sum_{{\bf q}\; {\rm odd}}\!\!
\cot \bigl(\pi({\bf p}-{\bf q})/2N\bigr)\braket{\bf
q}{\psi_{\ell-1}}  + i\sqrt N \Prob^{\frac12}(\ell-1)\Bigr|
}
and accordingly 
\numEq{3.12}{
\Prob(\ell) = \biggl[
\frac1N \sum_{{\bf p}\; {\rm even}}
\Bigl\{\Prob(\ell-1) + 
\frac1N \Bigl(\sum_{{\bf q}\; {\rm odd}}\!\!
\cot \frac{\pi({\bf p}-{\bf q})}{2N}\braket{\bf
q}{\psi_{\ell-1}}
\Bigr)^2\Bigr\}^{\frac12}
\biggr]^2\ . }
This formula shows that $\Prob(\ell)\ge\Prob(\ell-1)$.
Furthermore, if $\braket{\bf q}{\psi_{\ell-1}}=1/\sqrt N$ for all
$\bf q$ odd (which is equivalent to $\Prob(\ell-1)=1$) then by
\refEq{3.11} we have $\braket{\bf p}{\psi_\ell}=1/\sqrt N$ for
all $\bf p$ even and $\Prob(\ell)=1$. The greedy algorithm tends
toward this fixed point. 

\begin{table}[htb]
 \def\hf{\hfil\null}\def\!{\llap{\phantom{0}}}
\caption{Probability of success of the greedy algorithm, stopping after $k$
quantum queries.}\label{tbl:1}
\begin{center} 
\begin{tabular}{r@{\quad\enspace}|*6{r}}
$N$\hf &\llap{$k$} $=1$\hf & 2\hf & 3\hf & 4\hf & 5\hf &
6\hf\\
\hline
   64 &\quad \!.2036 & \!.6495 & \!.9615 & \!.9997 & 1.000\!
& 1.000\!
\\
 256 & \!.0788 & \!.3886 & \!.8221 & \!.9907 & \!.9999 & 1.000\! \\
1024 & \!.0282 & \!.2000 & \!.5981 & \!.9324 & \!.9983 & 1.000\! \\
2048 & \!.0165 & \!.1374 & \!.4818 & \!.8690 & \!.9939 & \!.9997 \\
4096 & \!.0096 & \!.0922 & \!.3755 & \!.7834 & \!.9819 & \!.9992 
\end{tabular}
\end{center}
\small Numbers are given to 4 significant figures, so the
1.000's do not mean exact performance.
\end{table}

We have some numerical results for the greedy algorithm, which
are presented in Table~\ref{tbl:1}. For
 these calculations we also
need the formulas analogous to \refEq{3.9} and \refEq{3.10} for
$\ell$~odd. Starting in the state $\ket{\psi_0} = \ket{\bf 0}$ it is
then straightforward to calculate $\braket{\bf p}{\psi_\ell}$ and
the associated probability of success. Clearly the greedy quantum
algorithm does much better than the best classical algorithm, which
has a probability of success of~$2^k/N$.

\section{Exact Algorithms}\label{sec:4}

An exactly successful $k$-query algorithm is a choice of $V_1,
V_2,\dots,V_k$ for which \refEq{16} holds. In this section, we
recast this condition in a form that allows us to determine, in
certain cases, if such a choice of $V$'s exists. 

For any $k$-query algorithm, successful or not, we define, as
before,
\begin{gather}\label{eq:4.1}
\ket{\psi_0} = \ket{\bf0}
\\
\label{eq:4.2}
\ket{\psi_\ell} = V_\ell\hat F_0 V_{\ell-1}\cdots V_1 \hat F_0
\ket{\bf0}
\end{gather}
where $1\le\ell\le k$. 
The form \refEq{20} for each $V_\ell$ implies by~\refEq{4.2}
\numEq{4.3}{
\Bigl| \braket{\bf p}{\psi_\ell} \Bigr| =
\Bigl| \bra{\bf p} \hat F_0  \ket{\psi_{\ell-1}} \Bigr| \ .
}
Conversely, given any sequence $\ket{\psi_0} ,\ket{\psi_1}
,\dots,\ket{\psi_k} $ satisfying \refEq{4.1} and
\refEq{4.3} there is a sequence $V_1,V_2,\dots,V_k$ of the form
\refEq{20} such that \refEq{4.2} holds with
\numEq{4.4new}{
e^{i\alpha_\ell({\bf p})} = 
\Cases{\displaystyle{
\frac{\braket{\bf p}{\psi_\ell}}{\bra{\bf p} \hat F_0 
\ket{\psi_{\ell-1}}}} & \quad \mbox{${\bf p}+\ell$ even}\\[2.5ex]
 1\hfil &  \quad \mbox{${\bf p}+\ell$ odd}} 
}
where the choice of $1$ for ${\bf p} + \ell$~odd is arbitrary.

If $\ket{\psi_\ell}$ satisfy
\refEq{4.2} and \refEq{4.1} [or equivalently \refEq{4.3} and
\refEq{4.1}] then as before  $\ket{\psi_\ell}$ is a superposition of
momentum basis states with $\bf p$~even for $\ell$~even and $\bf
p$~odd for $\ell$~odd. The corresponding statement in the
$x$~basis is
\numEq{4.4}{
\braket{x+N}{\psi_\ell} = (-1)^\ell \braket{x}{\psi_\ell}\ .
}
Using \refEq{4.4} and \refEq{18} we have
\numEq{4.5}{
\braket{\bf p}{\psi_\ell} = \Cases{\displaystyle{
\sqrt{\frac2N} \sum_{x=0}^{N-1}  \braket{x}{\psi_\ell}
e^{-i{\bf p}x\pi/N}} & \quad \mbox{${\bf p}+\ell$ even}\\
 0\hfil &  \quad \mbox{${\bf p}+\ell$ odd}}
}
and with \refEq{4} and \refEq{5} we have
\numEq{4.6}{
\bra{\bf p}\hat F_0 \ket{\psi_{\ell-1}} = \Cases{\displaystyle{
\sqrt{\frac2N} \sum_{x=0}^{N-1}  \braket{x}{\psi_{\ell-1}}
e^{-i{\bf p}x\pi/N}} & \quad \mbox{${\bf p}+\ell$ even}\\
 0\hfil  &  \quad \mbox{${\bf p}+\ell$ odd.}}
}
Thus \refEq{4.3} can be reformulated as
\numEq{4.7}{
\Bigl| \sum_{x=0}^{N-1} \braket{x}{\psi_\ell} z^{-x} \Bigr| =
\Bigl| \sum_{x=0}^{N-1} \braket{x}{\psi_{\ell-1}}
z^{-x}
\Bigr|\quad\mbox{at $z^N = (-1)^\ell$.}
 }
Define polynomials of degree $N-1$ in the complex variable~$z$ by 
\numEq{4.8}{
P_\ell(z)=\sqrt2 \sum_{x=0}^{N-1} \braket{x}{\psi_\ell}
z^{N-1-x}\ .
}
In terms of polynomials \refEq{4.8} the condition \refEq{4.7} is 
\numEq{4.9}{
\Bigl| P_\ell(z)\Bigr|  = \Bigl| P_{\ell-1} (z)\Bigr|\quad\mbox{at $z^N =
(-1)^\ell$.} }
 and 
\begin{align}\label{eq:4.10} 
P_0(z) &=\sqrt2 \sum_{x=0}^{N-1} \braket{x}{\bf 0}
z^{N-1-x} \\
&= \frac1{\sqrt N} (z^{N-1} + z^{N-2} + \dots + 1)\nonumber
\end{align}
Any sequence of degree $N-1$ polynomials, 
$P_0, P_1, P_2, \dots, P_k$ satisfying \refEq{4.9} and \refEq{4.10}
corresponds to a $k$-query algorithm. For a $k$-query algorithm
to be exactly successful we require, by \refEq{16}, that 
\begin{align}\label{4.11}
 \ket{\psi_k} &=  \frac1{\sqrt2} (\ket 0 + \ket N)  \quad
\mbox{when $k$ is even}\\
\noalign{\noindent or}
\ket{\psi_k} &=\frac1{\sqrt2} (\ket 0 - \ket N)  \quad
\mbox{when $k$ is odd}\nonumber
\end{align}
or equivalently
\numEq{4.12}{
P_k(z) = z^{N-1}\ .
}
To summarize, an exactly successful $k$-query
translationally invariant algorithm exists if and only if a sequence
of degree~$N-1$ polynomials $P_0, P_1, \dots, P_k$ exists that
satisfies 
\refEq{4.9}, with $P_0$ given by~\refEq{4.10} and  $P_k$ given
by~\refEq{4.12}.

Now define $Q_\ell$ by 
\numEq{4.13}{
Q_\ell(z) = P_\ell(z) \Bigl[ P_\ell \Bigl(\frac1{z^*}\Bigr)\Bigr]^*
}
and its coefficients $q_{\ell r}$ by 
\numEq{4.14}{
Q_\ell(z) = \sum_{r=-(N-1)}^{N-1}\!\! q_{\ell r} z^r\ .
}
Note that
\numEq{4.15}{
Q_\ell(z) =  \Bigl[ Q_\ell \Bigl(\frac1{z^*}\Bigr)\Bigr]^*\ , 
  \mbox{ that is, $q_{\ell r} =  q_{\ell, -r}^*$}
}
and 
\numEq{4.16}{
Q_\ell(z) \ge 0 \qquad\mbox{on $\left|z\right|=1$.} 
}
Now \refEq{4.9} is the same as 
\numEq{4.17}{
Q_\ell(z) = Q_{\ell-1} (z) \qquad\mbox{at $z^N = (-1)^\ell$} 
}
and \refEq{4.10} gives
\begin{align}\label{eq:4.18}
Q_0(z) &= \frac1N \bigl[ z^{N-1} + 2 z^{N-2}+\cdots + (N-1)z +N +
(N-1)z^{-1} +\cdots + z^{1-N}  \bigr]
\end{align}
and \refEq{4.12} gives 
\numEq{4.19}{
Q_k(z) = 1\ .
}
One of the reasons we have introduced the $Q$'s is that the condition
\refEq{4.17} will turn out to be more tractable than~\refEq{4.9}.

Given a sequence $Q_0, Q_1,\dots,Q_k$ defined by \refEq{4.13}
where the $P_\ell$'s satisfy \refEq{4.9}, \refEq{4.10},
and\refEq{4.12} it is immediate that \refEq{4.15}--\refEq{4.19} are
satisfied. We now establish the converse: given a sequence $Q_0,
Q_1,\dots,Q_k$ of the form \refEq{4.14} satisfying 
\refEq{4.15}--\refEq{4.19}, each $Q_\ell$ can be factored as
$P_\ell(z) [P_\ell(1/z^*)]^*$ where the polynomials $P_\ell$ satisfy 
\refEq{4.9}, \refEq{4.10}, and\refEq{4.12}.

The key ingredient in establishing this converse is to prove that any
$Q(z)$ of the form
\numEq{4.20}{
Q(z) = \sum_{r=-M}^M q_r z^r\ ,\quad\mbox{$q_M\neq0$}
}
satisfying 
\numEq{4.21}{
Q(z) =  \Bigl[ Q \Bigl(\frac1{z^*}\Bigr)\Bigr]^*\ , 
  \mbox{ that is, $q_{r} =  q_{-r}^*$}
}
and 
\numEq{4.22}{
Q(z) \ge 0\quad\mbox{on $\left| z\right| = 1$}
}
can be factored as 
\numEq{4.23}{
Q(z) =  P(z)\Bigl[ P \Bigl(\frac1{z^*}\Bigr)\Bigr]^*
}
for some polynomial $P$ of degree~$M$. 

Proof: $z^M Q(z)$ is a polynomial of degree~$2M$. Because of
\refEq{4.21} its zeros occur in pairs, $z=ae^{i\alpha}$ and
$z=\frac1ae^{i\alpha}$ ($a$ real and positive). The only
exception might be a zero on $\left| z\right| = 1$ but \refEq{4.22}
implies that such zeros have \emph{even} multiplicity. Thus we
can factor 
\numEq{4.24}{
z^M Q(z) = C \prod_{t=1}^M (z-a_t e^{i\alpha_t}) \Bigl(z-\frac1{a_t}
e^{i\alpha_t}\Bigr)  
}
or 
\numEq{4.25}{
 Q(z) = D \prod_{t=1}^M (z-a_t e^{i\alpha_t}) \Bigl(\frac1z -a_t
e^{-i\alpha_t}\Bigr)  \ .
}
Now \refEq{4.21} shows that $D=D^*$ and \refEq{4.22} shows that
$D>0$. Now take 
\numEq{4.26}{
 P(z) = \sqrt D \prod_{t=1}^M (z-a_t e^{i\alpha_t})
}
establishing~\refEq{4.23}.

Having established~\refEq{4.23} for each $Q_\ell$ obeying \refEq{4.15}
and \refEq{4.16} it then follows immediately that the corresponding
$P_\ell$'s obey \refEq{4.9} if the $Q_\ell$'s obey~\refEq{4.17}. We have
thus shown that the existence of a sequence $Q_0,
Q_1,\dots,Q_k$ satisfying \refEq{4.14}--\refEq{4.19} is equivalent to the
existence of an exactly successful $k$-query translationally invariant
algorithm. Our goal is now to try to determine for which values of~$N$
and~$k$ such a sequence exists.

Formula \refEq{4.15} implies that
\numEq{4.27}{
Q_\ell  (e^{i\theta}) = \sum_{r=0}^{N-1} C_{\ell r} \cos r\theta + 
 \sum_{r=1}^{N-1} S_{\ell r} \sin r\theta
}
where $ C_{\ell r}$ and  $S_{\ell r}$ are real. The matching condition
\refEq{4.17}, $Q_\ell (z) = Q_{\ell-1}(z)$ at $z^N=(-1)^\ell$, in terms of
\refEq{4.27} is 
\numEq{4.28}{
\sum_{r=0}^{N-1} (C_{\ell r}-C_{\ell-1, r} )\cos\Bigl( r\frac{2\pi
m}N\Bigr) +  \sum_{r=1}^{N-1} (S_{\ell r}-S_{\ell-1, r}) \sin
\Bigl(r\frac{2\pi m}N\Bigr) = 0
}
for $m=0,1,\dots,N-1$ in the case that $\ell$ is even. By taking the
sum and difference of \refEq{4.28} with~$m$ and $m$~replaced by 
$N-m$  we see that  the matching conditions on the $C_{\ell r}$
decouple from the matching conditions on the
$S_{\ell r}$ (and similarly for $\ell$~odd). Furthermore,
$Q_\ell(e^{i\theta})\ge 0$ along with $Q_\ell(e^{-i\theta})\ge 0$, which
follow from \refEq{4.16}, combine to give  
\numEq{4.29}{
\sum_{r=0}^{N-1} C_{\ell r} \cos r\theta \ge 
\Bigl| \sum_{r=1}^{N-1} S_{\ell r} \sin r\theta\Bigr| \ .
}
By \refEq{4.18} and \refEq{4.19} there are no $\sin r\theta$ terms in
$Q_0$ and $Q_k$, that is, $S_{0r} = S_{kr} = 0$. Thus we see that without
loss of generality we can set all $S_{\ell r}=0$ in~\refEq{4.27} while
attempting to determine if a sequence $Q_0,
Q_1,\dots,Q_k$ obeying \refEq{4.14}--\refEq{4.19} exists. 

Now by \refEq{4.8} and \refEq{4.13}, 
\numEq{4.30}{
Q_\ell (z) = 2 \sum_{x,y=0}^{N-1} \braket{\psi_\ell}y \braket
x{\psi_\ell} z^{y-x}
 }
and we see that the $z^0$ term is $\displaystyle 2\sum_{x=0}^{N-1} 
\bigl| \braket x{\psi_\ell}\bigr|^2$, which is~1 by~\refEq{4.4}. This
implies that $C_{\ell 0}$ in \refEq{4.27} is~1. Now decompose 
\refEq{4.27} as
\numEq{4.31}{
Q_\ell  (e^{i\theta}) = 1+ A_\ell(\theta) + B_\ell(\theta)
}
where
\begin{align}
 A_\ell(\theta) &= \sum_{r=1}^{N-1} a_{\ell r} \cos r\theta\ ,\quad
a_{\ell r} = a_{\ell, N- r} \label{eq:4.32}\\[1ex]
 B_\ell(\theta) &= \sum_{r=1}^{N-1} b_{\ell r} \cos r\theta\ ,\quad
b_{\ell r} = -b_{\ell, N- r}\ . \label{eq:4.33}
\end{align}
Because $ A_\ell(\theta) =0$ when $e^{iN\theta}=-1$ and 
$B_\ell(\theta) =0$ when $e^{iN\theta}=1$, the matching conditions
\refEq{4.17} become
\begin{align}
 B_1 &= B_0 \label{eq:4.34}\\
 A_2 &= A_1\nonumber\\
 B_3 &= B_2 \nonumber\\
 A_4 &= A_3\nonumber\\
 &\ \, \vdots \nonumber
\end{align}
From \refEq{4.18} we have
\numEq{4.35}{
Q_0 = 1 +\frac2N\bigl[
(N-1)\cos\theta + (N-2)\cos2\theta + \cdots +
\cos(N-1)\theta
\bigr]
}
and by \refEq{4.31}--\refEq{4.33}
\numEq{4.36}{
A_0 = 
\cos\theta + \cos2\theta + \cdots +
\cos(N-1)\theta
}
 and $B_0$ is given as
\numEq{4.37}{
B_0 =
\Bigl(1-\frac2N\Bigr)\cos\theta +
\Bigl(1-\frac4N\Bigr)\cos2\theta +
\cdots -
\Bigl(1-\frac2N\Bigr)\cos(N-1)\theta 
\ . }
By \refEq{4.19}
\numEq{4.38}{
A_k = B_k =0\ .
}

Finally, we can state the equivalence that we actually use.
The existence of an exactly successful $k$-query translationally
invariant algorithm is equivalent to the existence of a sequence of
functions $A_0, B_0, A_1, B_1$, $\dots, A_k, B_k$ of the form
\refEq{4.32} and \refEq{4.33} with $A_0, B_0$ given by
\refEq{4.36}, \refEq{4.37} and   $A_k$, 
$B_k$ given by \refEq{4.38} with the matching conditions
\refEq{4.34} and the positivity condition
\numEq{4.39}{
1 + A_\ell(\theta) + B_\ell (\theta) \ge 0
}
for $0\le\theta\le\pi$ and $0\le\ell\le k$. 

We now apply the machinery developed above to the 2-query case to
see for which $N$ an exactly successful translationally invariant
algorithm exists. The 2-query algorithm corresponds to the sequence 
$A_0, B_0, A_1, B_1, A_2, B_2$, with $A_2=B_2=0$ by \refEq{4.38}, 
$A_0$ and $B_0$ given by \refEq{4.36} and \refEq{4.37}, and
$B_1=B_0$ and $A_1=A_2=0$ by the matching conditions~\refEq{4.34}.
The condition \refEq{4.39} for $\ell=1$ becomes
\numEq{4.40}{
1 + B_0(\theta) \ge 0 
}
with $B_0$ given by \refEq{4.37}.
Numerical examination of \refEq{4.40} shows that this inequality
holds for $N\le6$ and we have shown that it does not for~$N\ge7$. Later
in this section we will explicitly show the $k=2$, $N=6$ algorithm. 

The $k=2$ case was particularly straightforward because the matching
conditions left no freedom to choose the $A$'s and $B$'s. For $k=3$ the
matching conditions leave a single undetermined function~$A_1$. The
two constraints that must be satisfied are \refEq{4.39} for $\ell=1$ and
$\ell=2$, that is
\begin{align}\label{eq:4.42}
1 + A_1(\theta) + B_0(\theta) &\ge 0\\
\mbox{\noindent
and} \hspace*{0.4\hsize}1 + A_1(\theta)&\ge
0\hspace*{0.4\hsize}
\nonumber
\end{align}
where $B_0$ is given by \refEq{4.37} 
and $A_1$ is of the form \refEq{4.32}. A 3-query
translationally invariant algorithm exists for a given~$N$ if and only if
such an $A_1$ can be found. By~\refEq{4.32}, $N/2$ (for $N$ even) real
parameters are needed to specify~$A_1$. By numerically searching we
have been able to find an $A_1$ that satisfies \refEq{4.42} for $N=52$.
This search was done on a laptop without heroic effort and we are 
 not claiming that 52 is best possible. 

We have shown that the existence of a sequence $Q_0,
Q_1,\dots,Q_k$ satisfying \refEq{4.14}--\refEq{4.19} implies the
existence of an exactly successful $k$-query translationally invariant
algorithm. Now we show explicitly how a given sequence $Q_0,
Q_1,\dots,Q_k$ determines the sequence of unitary operators $V_1,
V_2,\dots,V_k$ that comprise the actual algorithm via~\refEq{17}. 

First each $Q_\ell$ is factored as in \refEq{4.25}, and \refEq{4.26} is
used to find each~$P_\ell$. Now \refEq{4.8} is used to find $\braket
x{\psi_\ell}$ for $x=0,1,\dots,N-1$, and together with \refEq{4.4} yields
all the  $\braket x{\psi_\ell}$. Next we use \refEq{18} to obtain
$\braket{\bf p}{\psi_\ell}$ from $\braket x{\psi_\ell}$.

Determining the $V_\ell$'s means determining the phases
$\alpha_\ell({\bf p})$ for each $\bf p$. To use \refEq{4.4new}, we
need $\bra{\bf p} \hat F_0 \ket{\psi_{\ell-1}}$, which can be found
 from $\ket{\psi_{\ell-1}}$ by inserting a complete set of $\ket x$
states,
\begin{align}
\bra{\bf p} \hat F_0 \ket{\psi_{\ell-1}} &= 
\sum_{x=0}^{2N-1}\bra{\bf p} \hat F_0 \ket x
\braket x{\psi_{\ell-1}} \label{eq:44} \\
 &= \sum_{x=0}^{2N-1} \braket{\bf p} x   F_0(x) 
\braket x{\psi_{\ell-1}} \ . \nonumber
\end{align}

For $k=2$ and $N=6$ we numerically carried out the program just
outlined. The sequence is $Q_0, Q_1, Q_2$ with $Q_0$ and $Q_2$
fixed. As before, to get $Q_1$ we set $S_{\ell r}=0$ in \refEq{4.27}.
This means that $Q_1(e^{i\theta}) = 1 + B_0(\theta)$. To obtain
$Q_1(z)$ we go to \refEq{4.37} with $N=6$ and set $\cos
r\theta=(z^r +z^{-r})/2$. We then numerically factor the
10-th-degree polynomial $z^5 Q_1(z)$ and continue following the
procedure given above to obtain $\alpha_1({\bf p})$ and
$\alpha_2({\bf p})$. We convert to the  $\ket x$ basis, where
translation invariance means 
\numEq{4.44}{
\bra x V_\ell \ket y = \bra{x-y} V_\ell \ket 0
}
with $x-y < 0$  replaced by $x-y +12$. We find
\numEq{4.45}{\def\hf{\hfil\null}\def\!{\phantom{$-$}}
\begin{tabular}{r@{\quad\enspace}|cc}
$x$\hf &\quad $\bra{x} V_1 \ket 0$ &\quad $\bra{x} V_2 \ket
0$\\
\hline
  0 & \!.7572 & \!.9122\\
  1 & $-$.3473 &  $-$.2022\\ 
  2 & $-$.0034 &  $-$.0380 \\
  3 & $-$.0640 & \!.0736\\
  4 & $-$.1367 & \!.1258\\
  5 & $-$.2011 & \!.1286\\
  6 & \!.2428 & $-$.0878\\
  7 & \!.3473 & $-$.2022\\
  8 & \!.0034 & $-$.0380\\
  9 & \!.0640 & \!.0736\\
10 & \!.1367 & \!.1258\\
11 & \!.2011 & \!.1286
\end{tabular}
}
Note that $\bra x V_\ell \ket y$ are all real whenever the 
$S_{\ell r}$ in \refEq{4.27} are 0.

The interested reader can now check, using \refEq{5} for $\hat F_0$, which
is diagonal in the $x$ basis, that
\numEq{4.46}{
\frac1{\sqrt2} \bigl(\ket 0 + \ket 6\bigr) = V_2\hat F_0 V_1 \hat F_0\ket s
}
where $\displaystyle \ket s = \frac1{\sqrt{12}} \sum_{x=0}^{11} \ket x$.
By the translation invariance of $V_1$ and $V_2$ it then follows that
\numEq{4.47}{
\frac1{\sqrt2} \bigl(\ket j + \ket{j+6}\bigr) = V_2\hat F_j V_1 \hat F_j\ket s
} 
for $j=0,1,2,3,4,5$. The 6 states in \refEq{4.47} are an orthogonal set,
so \refEq{4.45} along with \refEq{4.44} is an explicit construction of an
exact algorithm for the $N=6$ insertion problem in 2 queries.

\section{Conclusion}\label{sec:5}

Symmetry plays a crucial role in  quantum physics. We have shown that there
are problems in which symmetry is useful in constructing quantum algorithms
that outperform the best classical algorithm. 

\subsubsection*{Acknowledgments}
One of us (E.F.) thanks Sean Robinson for useful discussions. We
thank Ron Rubin for stimulating discussions and Martin Stock
for \LaTeX-ing beyond the call of duty.


\begin{thebibliography}{99}

\bibitem{ref:1}
  E. Farhi, J. Goldstone, S. Gutmann, and M. Sipser,
     {\tt quant-ph/9812057}~.

\bibitem{ref:2}
 H. R\"ohrig,  {\tt quant-ph/9812061}~.

\bibitem{ref:3}
  L. K. Grover,  {\tt quant-ph/9605043}~.

\end{thebibliography}
\end{document}